\documentclass[letterpaper, 12pt]{article}[2000/05/19]
\usepackage[english]{babel}
\usepackage{amsfonts,amsmath,amssymb,amsthm,latexsym,amscd,mathrsfs}
\usepackage{ifthen,cite}
\usepackage[bookmarksnumbered=true]{hyperref}

\hypersetup{pdfpagetransition={Split}}

\newcommand{\evenhead}{Author \ name}
\newcommand{\oddhead}{Article \ name}
\newcommand{\theArticleName}{Article \ name}

\newcommand{\FirstPageHeading}[1]{\thispagestyle{empty}%
\noindent\raisebox{0pt}[0pt][0pt]{\makebox[\textwidth]{\protect\footnotesize \sf }}\par}

\newcommand{\ArticleName}[1]{\renewcommand{\theArticleName}{#1}\vspace{-2mm}\par\noindent {\LARGE\bf  #1\par}}
\newcommand{\Author}[1]{\vspace{5mm}\par\noindent {\Large  #1\par} \par\vspace{2mm}\par}
\newcommand{\Address}[1]{\vspace{2mm}\par\noindent {\it #1} \par}
\newcommand{\Email}[1]{\ifthenelse{\equal{#1}{}}{}{\par\noindent {\rm E-mail: }{\it  #1} \par}}
\newcommand{\URLaddress}[1]{\ifthenelse{\equal{#1}{}}{}{\par\noindent {\rm URL: }{\tt  #1} \par}}
\newcommand{\EmailD}[1]{\ifthenelse{\equal{#1}{}}{}{\par\noindent {$\phantom{\dag}$~\rm E-mail: }{\it  #1} \par}}
\newcommand{\URLaddressD}[1]{\ifthenelse{\equal{#1}{}}{}{\par\noindent {$\phantom{\dag}$~\rm URL: }{\tt  #1} \par}}

\newcommand{\Abstract}[1]{\vspace{6mm}\par\noindent\hspace*{10mm}
\parbox{140mm}{\small {\bf Abstract.} #1}\par}
\newcommand{\Keywords}[1]{\vspace{3mm}\par\noindent\hspace*{10mm}
\parbox{140mm}{\small {\bf Key words:} \rm #1}\par}
\newcommand{\Classification}[1]{\vspace{3mm}\par\noindent\hspace*{10mm}
\parbox{140mm}{\small {\it 2000 Mathematics Subject Classification:} \rm #1}\vspace{3mm}\par}
\newcommand{\ShortArticleName}[1]{\renewcommand{\oddhead}{#1}}
\newcommand{\AuthorNameForHeading}[1]{\renewcommand{\evenhead}{#1}}

\long\def\@makecaption#1#2{
  \sbox\@tempboxa{\small \textbf{#1.}\ \ #2}%
  \ifdim \wd\@tempboxa >\hsize
    {\small \textbf{#1.}\ \ #2}\par \else
    \global \@minipagefalse
    \hb@xt@\hsize{\hfil\box\@tempboxa\hfil}%
  \fi \vskip\belowcaptionskip}

\def\numberwithin#1#2{\@ifundefined{c@#1}{\@nocounterr{#1}}{%
  \@ifundefined{c@#2}{\@nocnterr{#2}}{%
  \@addtoreset{#1}{#2}%
  \toks@\@xp\@xp\@xp{\csname the#1\endcsname}%
  \@xp\xdef\csname the#1\endcsname
    {\@xp\@nx\csname the#2\endcsname.\the\toks@}}}}
\def\E^#1{{\buildrel #1 \over\vee}}
\newtheorem{theorem}{Theorem}

\newtheorem{corollary}{Corollary}

{\theoremstyle{definition}

}

\begin{document}

\FirstPageHeading{V.I. Gerasimenko}

\ShortArticleName{Quantum kinetic equations with correlations}

\AuthorNameForHeading{V.I. Gerasimenko}

\ArticleName{Mean Field Asymptotic Behavior of \\Quantum Particles with Initial Correlations}

\Author{V.I. Gerasimenko$^\ast$\footnote{E-mail: \emph{gerasym@imath.kiev.ua}}}

\Address{$^\ast$Institute of Mathematics of NAS of Ukraine,\\
    \hspace*{1mm}3, Tereshchenkivs'ka Str.,\\
    \hspace*{1mm}01601, Kyiv-4, Ukraine}

\bigskip

\Abstract{In the paper we consider the problem of the rigorous description of the kinetic evolution
in the presence of initial correlations of quantum large particle systems. One of the developed
approaches consists in the description of the evolution of quantum many-particle systems within
the framework of marginal observables in mean field scaling limit. Another method based on the
possibility to describe the evolution of states within the framework of a one-particle marginal
density operator governed by the generalized quantum kinetic equation in case of initial states
specified by a one-particle marginal density operator and correlation operators.}

\bigskip

\vspace{2pc}
\Classification{35Q20; 47J35.}

\makeatletter
\renewcommand{\@evenhead}{
\hspace*{-3pt}\raisebox{-7pt}[\headheight][0pt]{\vbox{\hbox to \textwidth {\thepage \hfil \evenhead}\vskip4pt \hrule}}}
\renewcommand{\@oddhead}{
\hspace*{-3pt}\raisebox{-7pt}[\headheight][0pt]{\vbox{\hbox to \textwidth {\oddhead \hfil \thepage}\vskip4pt\hrule}}}
\renewcommand{\@evenfoot}{}
\renewcommand{\@oddfoot}{}
\makeatother

\newpage
\vphantom{math}

\protect\tableofcontents
\vspace{0.7cm}

\section{Introduction}
As is known the collective behavior of quantum many-particle systems can be effectively
described within the framework of a one-particle marginal density operator governed by
the kinetic equation in a suitable scaling limit of underlying dynamics. At present the
considerable advances in the rigorous derivation of the quantum kinetic equations in
the mean (self-consistent) field scaling limit is observed \cite{Go13}-\cite{ESch}.
In particular, the nonlinear Schr\"{o}dinger equation \cite{PP09}-\cite{KSS} and the
Gross--Pitaevskii equation \cite{ESchY2}-\cite{Ch} was justified.

The conventional approach to this problem is based on the consideration of an asymptotic
behavior of a solution of the quantum BBGKY hierarchy for marginal density operators
constructed within the framework of the theory of perturbations in case of initial
data specified by one-particle marginal density operators without correlations, i.e.
such that satisfy a chaos condition \cite{Sp91},\cite{CGP97}.

In paper \cite{GT} it was developed more general method of the derivation of the quantum
kinetic equations. By means of a non-perturbative solution of the quantum BBGKY hierarchy
constructed in \cite{GerS} it was established that, if initial data is completely specified
by a one-particle marginal density operator, then all possible states of many-particle
systems at arbitrary moment of time can be described within the framework of a one-particle
density operator governed by the generalized quantum kinetic equation (see also \cite{G12}).
Then the actual quantum kinetic equations can be derived from the generalized quantum kinetic
equation in appropriate scaling limits, for example, in a mean field limit \cite{GTrmp}.

Another approach to the description of the many-particle evolution is given within the
framework of marginal observables governed by the dual quantum BBGKY hierarchy \cite{BG}.
In paper \cite{G11} a rigorous formalism for the description of the kinetic evolution of
observables of quantum particles in a mean field scaling limit was developed.

In this paper we consider the problem of the rigorous description of the kinetic evolution
in the presence of initial correlations of quantum particles. Such initial states are typical
for the condensed states of quantum gases in contrast to the gaseous state. For example, the
equilibrium state of the Bose condensate satisfies the weakening of correlation condition
specified by correlations of the condensed state \cite{BogLect}. One more example is the
influence of initial correlations on ultrafast relaxation processes in plasmas \cite{SKB},\cite{KB}.

Thus, our goal consists in the rigorous derivation of the quantum kinetic equations
in the presence of initial correlations of quantum large particle systems.

We outline the structure of the paper.
In section 2, we establish the mean field asymptotic behavior of marginal observables governed
by the dual quantum BBGKY hierarchy. The limit dynamics is described by the set of recurrence
evolution equations, namely by the dual quantum Vlasov hierarchy. Furthermore, the links of the
dual quantum Vlasov hierarchy for the limit marginal observables and the quantum Vlasov-type
kinetic equation with initial correlations are established.
In section 3, we consider the relationships of dynamics described by marginal observables and
within the framework of a one-particle marginal density operator governed by the generalized
quantum kinetic equation in the presence of initial correlations.
In section 4, we develop one more approach to the description of the quantum kinetic evolution
with initial correlations in the mean field limit. We prove that a solution of the generalized
quantum kinetic equation with initial correlations is governed by the quantum Vlasov-type equation
with initial correlations. The property of the propagation of initial correlations is also
established.
Finally, in section 5, we conclude with some perspectives for future research.


\section{The kinetic evolution within the framework of marginal observables}

The kinetic evolution of many-particle systems can be described within the framework
of observables. We consider this problem on an example of the mean field asymptotic
behavior of a non-perturbative solution of the dual quantum BBGKY hierarchy for
marginal observables. Moreover, we establish the links of the dual quantum Vlasov
hierarchy for the limit marginal observables with the quantum Vlasov-type kinetic
equation in the presence of initial correlations.

\subsection{Many-particle dynamics of observables}
We shall consider a quantum system of a non-fixed (i.e. arbitrary but finite) number of identical
(spinless) particles obeying Maxwell--Boltzmann statistics in the space $\mathbb{R}^{3}$.
We will use units where $h={2\pi\hbar}=1$ is a Planck constant, and $m=1$ is the mass of particles.

Let the space $\mathcal{H}$ be a one-particle Hilbert space, then the $n$-particle space
$\mathcal{H}_n=\mathcal{H}^{\otimes n}$ is a tensor product of $n$ Hilbert spaces $\mathcal{H}$.
We adopt the usual convention that $\mathcal{H}^{\otimes 0}=\mathbb{C}$. The Fock space over the
Hilbert space $\mathcal{H}$ we denote by
$\mathcal{F}_{\mathcal{H}}={\bigoplus\limits}_{n=0}^{\infty}\mathcal{H}_{n}$.

The Hamiltonian $H_{n}$ of a system of $n$ particles is a self-adjoint operator with the domain
$\mathcal{D}(H_{n})\subset\mathcal{H}_{n}$
\begin{eqnarray}\label{H}
    &&H_{n}=\sum\limits_{i=1}^{n}K(i)+\epsilon\sum\limits_{i_{1}<i_{2}=1}^{n}\Phi(i_{1},i_{2}),
\end{eqnarray}
where $K(i)$ is the operator of a kinetic energy of the $i$ particle, $\Phi(i_{1},i_{2})$ is the
operator of a two-body interaction potential and $\epsilon>0$ is a scaling parameter. The operator
$K(i)$ acts on functions $\psi_n$, that belong to the subspace
$L^{2}_{0}(\mathbb{R}^{3n})\subset\mathcal{D}(H_n)\subset L^{2}(\mathbb{R}^{3n})$ of infinitely
differentiable functions with compact supports, according to the formula:
$K(i)\psi_n=-\frac{1}{2}\Delta_{q_i}\psi_n$. Correspondingly, we have:
$\Phi(i_{1},i_{2})\psi_{n}=\Phi(q_{i_{1}},q_{i_{2}})\psi_{n}$, and we assume that the function
$\Phi(q_{i_{1}},q_{i_{2}})$ is symmetric with respect to permutations of its arguments,
translation-invariant and bounded function.

Let a sequence $g=(g_{0},g_{1},\ldots,g_{n},\ldots)$ be an infinite sequence of self-adjoint
bounded operators $g_{n}$ defined on the Fock space $\mathcal{F}_{\mathcal{H}}$. An operator
$g_{n}$ defined on the $n$-particle Hilbert space $\mathcal{H}_{n}=\mathcal{H}^{\otimes n}$
will be also denoted by the symbol $g_{n}(1,\ldots,n)$.
Let the space $\mathfrak{L}(\mathcal{F}_\mathcal{H})$ be the space of sequences
$g=(g_{0},g_{1},\ldots,$ $g_{n},\ldots)$ of bounded operators $g_{n}$ defined on the Hilbert
space $\mathcal{H}_n$ that satisfy symmetry condition:
$g_{n}(1,\ldots,n)=g_{n}(i_1,\ldots,i_n)$, for arbitrary $(i_{1},\ldots,i_{n})\in (1,\ldots,n)$,
equipped with the operator norm $\|.\|_{\mathfrak{L}(\mathcal{H}_{n})}$. We will also consider
a more general space $\mathfrak{L}_{\gamma}(\mathcal{F}_\mathcal{H})$ with the norm
\begin{eqnarray*}
   &&\big\|g\big\|_{\mathfrak{L}_{\gamma} (\mathcal{F}_\mathcal{H})}\doteq
      \max\limits_{n\geq 0}\,\frac{\gamma^n}{n!}\,\big\|g_{n}\big\|_{\mathfrak{L}(\mathcal{H}_{n})},
\end{eqnarray*}
where $0<\gamma<1$. We denote by $\mathfrak{L}_{_{\gamma},0}(\mathcal{F}_\mathcal{H})
\subset\mathfrak{L}_{\gamma}(\mathcal{F}_\mathcal{H})$ the everywhere dense set in the space
$\mathfrak{L}_{\gamma}(\mathcal{F}_\mathcal{H})$ of finite sequences of degenerate operators
with infinitely differentiable kernels with compact supports.

For $g_n\in\mathfrak{L}(\mathcal{H}_{n})$ it is defined the one-parameter mapping
\begin{eqnarray}\label{grG}
    &&\mathbb{R}^1\ni t\mapsto\mathcal{G}_n(t)g_n\doteq e^{itH_{n}}g_n e^{-itH_{n}},
\end{eqnarray}
where the Hamilton operator $H_{n}$ has the structure \eqref{H}. On the space
$\mathfrak{L}(\mathcal{H}_{n})$ one-parameter mapping \eqref{grG} is an isometric $\ast$-weak
continuous group of operators. The infinitesimal generator $\mathcal{N}_{n}$ of this group of
operators is a closed operator for the $\ast$-weak topology, and on its domain of the definition
$\mathcal{D}(\mathcal{N}_{n})\subset\mathfrak{L}(\mathcal{H}_{n})$ it is defined in the sense of
the $\ast$-weak convergence of the space $\mathfrak{L}(\mathcal{H}_{n})$ by the operator
\begin{eqnarray}\label{infOper1}
    &&\mathrm{w^{\ast}-}\lim\limits_{t\rightarrow 0}\frac{1}{t}\big(\mathcal{G}_n(t)g_n-g_n \big)
    =-i\,(g_n H_n-H_n g_n)\doteq\mathcal{N}_n g_n,
\end{eqnarray}
where $H_{n}$ is the Hamiltonian \eqref{H} and the operator $\mathcal{N}_n g_n$ defined
on the domain $\mathcal{D}(H_n)\subset\mathcal{H}_n$ has the structure
\begin{eqnarray*}
    &&\mathcal{N}_n=\sum\limits_{j=1}^{n}\mathcal{N}(j)+
       \epsilon\sum\limits_{j_{1}<j_{2}=1}^{n}\mathcal{N}_{\mathrm{int}}(j_{1},j_{2}),
\end{eqnarray*}
where
\begin{eqnarray}\label{0}
   &&\mathcal{N}(j)g_n\doteq -i\,(g_n K(j)-K(j)g_n),
\end{eqnarray}
and
\begin{eqnarray}\label{int}
   &&\mathcal{N}_{\mathrm{int}}(j_{1},j_{2})g_n\doteq
     -i\,(g_n \Phi(j_{1},j_{2})-\Phi(j_{1},j_{2})g_n).
\end{eqnarray}

Therefore on the space $\mathfrak{L}(\mathcal{H}_{n})$ a unique solution of the Heisenberg
equation for observables of a $n$-particle system is determined by group \eqref{grG} \cite{G12}.

In what follows in this Section we shall hold abridged notations:
$Y\equiv(1,\ldots,s),\,X\equiv(j_1,\ldots,j_{n})\subset Y$, and $\{Y\setminus X\}$ is the set,
consisting of a single element $Y\setminus X=(1,\ldots,s)\setminus(j_1,\ldots,j_{n})$, thus,
the set $\{Y\setminus X\}$ is a connected subset of the set $Y$.

To describe the evolution within the framework of marginal observables \cite{BG} we introduce a notion
of the $(1+n)th$-order ($n\geq0$) cumulant of groups of operators \eqref{grG} as follows \cite{GerS}
\begin{eqnarray}\label{cumulant}
    &&\hskip-7mm \mathfrak{A}_{1+n}(t,\{Y\setminus X\},X)\doteq
       \sum\limits_{\mathrm{P}:\,(\{Y\setminus X\},\,X)={\bigcup}_i X_i}
       (-1)^{\mathrm{|P|}-1}({\mathrm{|P|}-1})!
       \prod_{X_i\subset \mathrm{P}}\mathcal{G}_{|\theta(X_i)|}(t,\theta(X_i)),
\end{eqnarray}
where the symbol ${\sum}_\mathrm{P}$ means the sum over all possible partitions $\mathrm{P}$
of the set $(\{Y\setminus X\},j_1,\ldots,j_{n})$ into $|\mathrm{P}|$ nonempty
mutually disjoint subsets $ X_i\subset(\{Y\setminus X\},X)$, and $\theta(\cdot)$ is the
declusterization mapping defined as follows: $\theta(\{Y\setminus X\},\,X)=Y$. For example,
\begin{eqnarray*}
    &&\mathfrak{A}_{1}(t,\{Y\})=\mathcal{G}_{s}(t,Y),\\
    &&\mathfrak{A}_{2}(t,\{Y\setminus (j)\},j)=\mathcal{G}_{s}(t,Y)
       -\mathcal{G}_{s-1}(t,Y\setminus(j))\mathcal{G}_{1}(t,j).\nonumber
\end{eqnarray*}

In terms of observables the evolution of quantum many-particle systems is described
by the sequence $B(t)=(B_0,B_{1}(t,1),\ldots,B_{s}(t,1,\ldots,s),\ldots)$ of marginal
observables (or $s$-particle observables) $B_{s}(t,1,\ldots,s),\,s\geq 1,$ determined
by the following expansions \cite{BG}:
\begin{eqnarray}\label{sdh}
   &&\hskip-7mm B_{s}(t,Y)=\sum_{n=0}^s\,\frac{1}{n!}\sum_{j_1\neq\ldots\neq j_{n}=1}^s
       \mathfrak{A}_{1+n}(t,\{Y\setminus X\},X)\,B_{s-n}^{0,\epsilon}(Y\setminus X),\quad s\geq1,
\end{eqnarray}
where $B(0)=(B_0,B_{1}^{0,\epsilon}(1),\ldots,B_{s}^{0,\epsilon}(1,\ldots,s),\ldots)\in
\mathfrak{L}_{\gamma}(\mathcal{F}_\mathcal{H})$ is a sequence of initial marginal observables,
and the generating operator $\mathfrak{A}_{1+n}(t)$ of expansion \eqref{sdh} is the
$(1+n)th$-order cumulant of groups of operators \eqref{grG} defined by expansion \eqref{cumulant}.
The simplest examples of marginal observables \eqref{sdh} are given by the expressions:
\begin{eqnarray*}
   &&B_{1}(t,1)=\mathfrak{A}_{1}(t,1)B_{1}^{0,\epsilon}(1),\\
   &&B_{2}(t,1,2)=\mathfrak{A}_{1}(t,\{1,2\})B_{2}^{0,\epsilon}(1,2)+
      \mathfrak{A}_{2}(t,1,2)(B_{1}^{0,\epsilon}(1)+B_{1}^{0,\epsilon}(2)).
\end{eqnarray*}

If $\gamma<e^{-1}$, for the sequence of operators \eqref{sdh} the following estimate is true
\begin{eqnarray*}
   &&\big\|B(t)\big\|_{\mathfrak{L}_{\gamma}(\mathcal{F}_\mathcal{H})}
        \leq e^2(1-\gamma e)^{-1}\big\|B(0)\big\|_{\mathfrak{L}_{\gamma}(\mathcal{F}_\mathcal{H})}.
\end{eqnarray*}

We note that a sequence of marginal observables \eqref{sdh} is the non-perturbative solution
of recurrence evolution equations known as the dual quantum BBGKY hierarchy \cite{BG}:
\begin{eqnarray*}
  \label{dh}
   &&\frac{\partial}{\partial t}B_{s}(t,Y)=\big(\sum\limits_{j=1}^{s}\mathcal{N}(j)+
      \sum\limits_{j_1<j_{2}=1}^{s}\mathcal{N}_{\mathrm{int}}(j_1,j_{2})\big)B_{s}(t,Y)+\\
   &&\hskip+21mm+\sum_{j_1\neq j_{2}=1}^s\mathcal{N}_{\mathrm{int}}(j_1,j_{2})
      B_{s-1}(t,Y\backslash (j_1)),\nonumber\\ \nonumber\\
  \label{dhi}
   &&B_{s}(t)|_{t=0}=B_{s}^0,\quad s\geq1.
\end{eqnarray*}
We adduce also the relationship of marginal observables governed by hierarchy (\ref{sdh})
and observables governed the Heisenberg equations \cite{G12}
\begin{eqnarray*}\label{mo}
      &&B_{s}(t,Y)\doteq\sum_{n=0}^s\,\frac{(-1)^n}{n!}\sum_{j_1\neq\ldots\neq j_{n}=1}^s
           (\mathcal{G}_{s-n}(t) A_{s-n}^0)(Y\setminus(j_1,\ldots,j_{n})), \quad s\geq 1,
\end{eqnarray*}
where the group $\mathcal{G}_{s-n}(t)$ is defined by formula (\ref{grG}) and the operators
$A_{s-n}^0,\,0\leq n\leq s$, are initial observables.

\subsection{A mean field asymptotic behavior of marginal observables}
A mean field asymptotic behavior of marginal observables (\ref{sdh}) is described by
the following theorem \cite{G11}.
\begin{theorem}\label{3.1}
Let for $B_{n}^{0,\epsilon}\in \mathfrak{L}(\mathcal{H}_{n}),\,n\geq1,$ in the sense
of the $\ast$-weak convergence on the space $\mathfrak{L}(\mathcal{H}_{s})$ it holds:
$\mathrm{w^{\ast}-}\lim_{\epsilon\rightarrow 0}(\epsilon^{-n}B_{n}^{0,\epsilon}-b_{n}^0)=0,$
then for arbitrary finite time interval there exists the mean field limit of marginal
observables (\ref{sdh}):
$\mathrm{w^{\ast}-}\lim_{\epsilon\rightarrow 0}(\epsilon^{-s}B_{s}(t)-b_{s}(t))=0,\,s\geq1,$
that are determined by the following expansions:
\begin{eqnarray}\label{Iterd}
   &&\hskip-9mm b_{s}(t,Y)=\sum\limits_{n=0}^{s-1}\,\int\limits_0^tdt_{1}\ldots\int\limits_0^{t_{n-1}}dt_{n}
      \prod\limits_{l_{1}\in Y}\mathcal{G}_{1}(t-t_{1},l_{1})
      \sum\limits_{i_{1}\neq j_{1}=1}^{s}\mathcal{N}_{\mathrm{int}}(i_{1},j_{1})
      \prod\limits_{l_{2}\in Y\setminus(j_{1})}\mathcal{G}_{1}(t_{1}-t_{2},l_{2})\ldots\\
   &&\prod\limits_{l_{n}\in Y\setminus(j_{1},\ldots,j_{n-1})}\mathcal{G}_{1}(t_{n-1}-t_{n},l_{n})
       \sum\limits^{s}_{\mbox{\scriptsize $\begin{array}{c}i_{n}\neq j_{n}=1,\\
       i_{n},j_{n}\neq (j_{1},\ldots,j_{n-1})\end{array}$}}\mathcal{N}_{\mathrm{int}}(i_{n},j_{n})\nonumber\\
   &&\times\prod\limits_{l_{n+1}\in Y\setminus(j_{1},\ldots,j_{n})}\mathcal{G}_{1}(t_{n},l_{n+1})
       b_{s-n}^0(Y\setminus (j_{1},\ldots,j_{n})),\nonumber
\end{eqnarray}
where the operator $\mathcal{N}_{\mathrm{int}}(i_{1},j_{2})$ is defined on
$g_n\in\mathfrak{L}(\mathcal{H}_{n})$ by formula \eqref{int}.
\end{theorem}

The proof of Theorem 1 is based on formulas for cumulants of asymptotically perturbed groups
of operators \eqref{grG}.

For arbitrary finite time interval the asymptotically perturbed group of operators \eqref{grG}
has the following scaling limit in the sense of the $\ast$-weak convergence on the space
$\mathfrak{L}(\mathcal{H}_{s})$:
\begin{eqnarray}\label{Kato}
    &&\mathrm{w^{\ast}-}\lim\limits_{\epsilon\rightarrow 0}\big(\mathcal{G}_s(t,Y)-
       \prod\limits_{j=1}^{s}\mathcal{G}_{1}(t,j)\big)g_s=0.
\end{eqnarray}
Taking into account analogs of the Duhamel equations for cumulants of asymptotically
perturbed groups of operators, in view of formula (\ref{Kato}) we have
\begin{eqnarray*}\label{apc}
   &&\hskip-8mm \mathrm{w^{\ast}-}\lim\limits_{\epsilon\rightarrow 0}
       \Big(\epsilon^{-n}\frac{1}{n!}\mathfrak{A}_{1+n}\big(t,\{Y\setminus X\},j_1,\ldots,j_{n}\big)-\\
   &&\hskip-8mm-\int\limits_0^tdt_{1}\ldots\int\limits_0^{t_{n-1}}dt_{n}
      \prod\limits_{l_{1}\in Y}\mathcal{G}_{1}(t-t_{1},l_{1})\sum\limits_{i_{1}\neq j_{1}=1}^{s}
      \mathcal{N}_{\mathrm{int}}(i_{1},j_{1})\prod\limits_{l_{2}\in Y\setminus(j_{1})}
      \mathcal{G}_{1}(t_{1}-t_{2},l_{2})\ldots\\
   &&\hskip-8mm \prod\limits_{l_{n}\in Y\setminus(j_{1},\ldots,j_{n-1})}
      \mathcal{G}_{1}(t_{n-1}-t_{n},l_{n})\sum\limits^{s}_{\mbox{\scriptsize $\begin{array}{c}i_{n}\neq j_{n}=1,\\
       i_{n},j_{n}\neq (j_{1},\ldots,j_{n-1})\end{array}$}}\hskip-2mm\mathcal{N}_{\mathrm{int}}(i_{n},j_{n})
       \prod\limits_{l_{n+1}\in Y \setminus(j_{1},\ldots,j_{n})}\mathcal{G}_{1}(t_{n},l_{n+1})\Big)g_{s-n}=0,
\end{eqnarray*}
where we used notations accepted in formula (\ref{Iterd}) and
$g_{s-n}\equiv g_{s-n}((1,\ldots,s)\setminus(j_{1},\ldots,j_{n})),\, n\geq1$.
As a result of this equality we establish the validity of Theorem 1 for expansion
(\ref{sdh}) of marginal observables.

If $b^0\in \mathfrak{L}_{\gamma}(\mathcal{F}_\mathcal{H})$, then the sequence
$b(t)=(b_0,b_1(t),\ldots,b_{s}(t),\ldots)$ of limit marginal observables (\ref{Iterd})
is a generalized global solution of the Cauchy problem of the dual quantum Vlasov hierarchy
\begin{eqnarray}\label{vdh}
  &&\hskip-12mm \frac{\partial}{\partial t}b_{s}(t,Y)=\sum\limits_{j=1}^{s}\mathcal{N}(j)\,b_{s}(t,Y)+
    \sum_{j_1\neq j_{2}=1}^s\mathcal{N}_{\mathrm{int}}(j_1,j_{2})\,b_{s-1}(t,Y\setminus(j_1)),\\
    \nonumber\\
 \label{vdhi}
   &&\hskip-12mm b_{s}(t)|_{t=0}=b_{s}^0,\quad s\geq1,
\end{eqnarray}
where the infinitesimal generator $\mathcal{N}(j)$ of the group of operators
$\mathcal{G}_{1}(t,j)$ of $j$ particle is defined on $g_n\in\mathfrak{L}_{0}(\mathcal{H}_{n})$ by
formular (\ref{0}).

It should be noted that equations set (\ref{vdh}) has the structure of recurrence evolution equations.
We give several examples of the evolution equations of the dual quantum Vlasov hierarchy (\ref{vdh}) in
terms of operator kernels of the limit marginal observables
\begin{eqnarray*}
    &&\hskip-9mm i\,\frac{\partial}{\partial t}b_{1}(t,q_1;q'_1)=-\frac{1}{2}(-\Delta_{q_1}+\Delta_{q'_1})b_{1}(t,q_1;q'_1),\\
    &&\hskip-9mm i\,\frac{\partial}{\partial t}b_{2}(t,q_1,q_2;q'_1,q'_2)=
       -\frac{1}{2}\sum\limits_{i=1}^2(-\Delta_{q_i}+\Delta_{q'_i})b_{2}(t,q_1,q_2;q'_1,q'_2)+\\
    &&+\big(\Phi(q'_1-q'_2)-\Phi(q_1-q_2)\big)\big(b_{1}(t,q_1;q'_1)+b_{1}(t,q_2;q'_2)\big).
\end{eqnarray*}

We consider the mean field limit of a particular case of marginal observables, namely the
additive-type marginal observables $B^{(1)}(0)=(0,B_{1}^{0,\epsilon}(1),0,\ldots)$. We remark
that the $k$-ary  marginal observables are represented by the sequence
$B^{(k)}(0)=\big(0,\ldots,0,B_{k}^{0,\epsilon}(1,\ldots,k),0,\ldots\big)$.
In case of additive-type marginal observables expansions (\ref{sdh}) take the following form:
\begin{eqnarray}\label{af}
    &&B_{s}^{(1)}(t,Y)=\mathfrak{A}_{s}(t)\,\sum_{j=1}^s\,B_{1}^{0,\epsilon}(j),\quad s\geq 1,
\end{eqnarray}
where the operator $\mathfrak{A}_{s}(t)$ is $s$-order cumulant (\ref{cumulant}) of groups of
operators (\ref{grG}).

\begin{corollary}
If for the additive-type marginal observable $B_{1}^{0,\epsilon}\in \mathfrak{L}(\mathcal{H})$ it holds
$\mathrm{w^{\ast}-}\lim_{\epsilon\rightarrow 0}(\epsilon^{-1} B_{1}^{0,\epsilon}-b_{1}^{0})=0,$
then, according to statement of Theorem 1, for additive-type marginal observable (\ref{af}) we have:
$\mathrm{w^{\ast}-}\lim_{\epsilon\rightarrow 0}(\epsilon^{-s}B_{s}^{(1),\epsilon}(t)-b_{s}^{(1)}(t))=0,\,s\geq1,$
where the limit additive-type marginal observable $b_{s}^{(1)}(t)$ is determined by a special case
of expansion (\ref{Iterd})
\begin{eqnarray}\label{itvad}
   &&\hskip-12mm b_{s}^{(1)}(t,Y)=\int\limits_0^tdt_{1}\ldots\int\limits_0^{t_{s-2}}dt_{s-1}
      \prod\limits_{l_{1}\in Y}\mathcal{G}_{1}(t-t_{1},l_{1})\sum\limits_{i_{1}\neq j_{1}=1}^{s}
      \mathcal{N}_{\mathrm{int}}(i_{1},j_{1})\\
   &&\times \prod\limits_{l_{2}\in Y\setminus(j_{1})}\mathcal{G}_{1}(t_{1}-t_{2},l_{2})\ldots
      \prod\limits_{l_{s-1}\in Y\setminus(j_{1},\ldots,j_{s-2})}\mathcal{G}_{1}(t_{s-2}-t_{s-1},l_{s-1})\nonumber\\
   &&\hskip-5mm \times\sum\limits^{s}_{\mbox{\scriptsize $\begin{array}{c}i_{s-1}\neq j_{s-1}=1,\\
       i_{s-1},j_{s-1}\neq (j_{1},\ldots,j_{s-2})\end{array}$}}\hskip-2mm\mathcal{N}_{\mathrm{int}}(i_{s-1},j_{s-1})
       \prod\limits_{l_{s}\in Y \setminus(j_{1},\ldots,j_{s-1})}\mathcal{G}_{1}(t_{s-1},l_{s})
       b_{1}^{0}(Y\setminus(j_{1},\ldots,j_{s-1})).\nonumber
\end{eqnarray}
\end{corollary}
We make several examples of expansions (\ref{itvad}) for the limit additive-type marginal observables
\begin{eqnarray*}
   &&\hskip-8mm b_{1}^{(1)}(t,1)=\mathcal{G}_{1}(t,1)\,b_{1}^{0}(1),\\
   &&\hskip-8mm b_{2}^{(1)}(t,1,2)=\int\limits_0^t dt_{1}\prod\limits_{i=1}^{2}\mathcal{G}_{1}(t-t_{1},i)
      \mathcal{N}_{\mathrm{int}}(1,2)\sum\limits_{j=1}^{2}\mathcal{G}_{1}(t_{1},j)\,b_{1}^{0}(j).
\end{eqnarray*}

Thus, for arbitrary initial states in the mean field scaling limit the kinetic evolution
of quantum many-particle systems is described in terms of limit marginal observables \eqref{Iterd}
governed by the dual quantum Vlasov hierarchy \eqref{vdh}.

\subsection{The derivation of the quantum Vlasov-type kinetic equation with initial correlations}
Furthermore, the relationships between the evolution of observables and the kinetic evolution
of states described in terms of a one-particle marginal density operator are discussed.

We shall consider initial states of a quantum many-particle system specified by the one-particle
(marginal) density operator $F_1^{0,\epsilon}\in\mathfrak{L}^{1}(\mathcal{H})$ in the presence of
correlations, i.e. initial state specified by the following sequence of marginal density operators
\begin{eqnarray}\label{ins}
   &&\hskip-8mm F^{c}=\big(1,F_1^{0,\epsilon}(1),g_{2}^{\epsilon}\prod_{i=1}^{2}F_1^{0,\epsilon}(i),\ldots,
        g_{n}^{\epsilon}\prod_{i=1}^{n}F_1^{0,\epsilon}(i),\ldots\big),
\end{eqnarray}
where the bounded operators
$g_{n}^{\epsilon}\equiv g_{n}^{\epsilon}(1,\ldots,n)\in\mathfrak{L}(\mathcal{H}_n),\,n\geq2$, are
specified the initial correlations. We remark that such assumption about initial states is intrinsic
for the kinetic description of a gas. On the other hand, initial data \eqref{ins} is typical for the
condensed states of quantum gases, for example, the equilibrium state of the Bose condensate satisfies
the weakening of correlation condition with the correlations which characterize the condensed state \cite{BogLect}.

We assume that for the initial one-particle (marginal) density operator
$F_{1}^{0,\epsilon}\in\mathfrak{L}^{1}(\mathcal{H})$ exists the mean field limit
$\lim_{\epsilon\rightarrow 0}\big\|\epsilon\,F_{1}^{0,\epsilon}-f_{1}^0\big\|_{\mathfrak{L}^{1}(\mathcal{H})}=0,$
and it holds: $\lim_{\epsilon\rightarrow 0}\big\|g_{n}^{\epsilon}-g_{n}\big\|_{\mathfrak{L}^{1}(\mathcal{H}_n)}=0,$
then in the mean field limit initial state is specified by the following sequence of limit operators
\begin{eqnarray}\label{lins}
   &&\hskip-8mm f^{c}=\big(1,f_1^{0}(1),g_{2}\prod_{i=1}^{2}f_1^{0}(i),\ldots,
        g_{n}\prod_{i=1}^{n}f_1^{0}(i),\ldots\big).
\end{eqnarray}

We note that in case of initial states specified by sequence \eqref{lins}
the average values (mean values) of limit marginal observables \eqref{Iterd} are determined by
the following positive continuous linear functional \cite{G12}
\begin{eqnarray}\label{averageb}
     &&\hskip-8mm \big(b(t),f^{c}\big)\doteq\sum\limits_{n=0}^{\infty}\frac{1}{n!}
         \,\mathrm{Tr}_{1,\ldots,n}\,b_{n}(t,1,...,n)\,g_{n}(1,...,n)\prod_{i=1}^{n}f_1^{0}(i).
\end{eqnarray}
For $b(t)\in \mathfrak{L}_{\gamma}(\mathcal{F}_\mathcal{H})$ and
$f_1^0\in \mathfrak{L}^{1}(\mathcal{H})$, functional \eqref{averageb} exists under the condition that
$\|f_1^0\|_{\mathfrak{L}^{1}(\mathcal{H})}<\gamma$.

We consider relationships of the constructed mean field asymptotic behavior of marginal
observables with the quantum Vlasov-type kinetic equation in case of initial states \eqref{lins}.

For the limit additive-type marginal observables (\ref{itvad}) the following equality is true
\begin{eqnarray*}\label{avmar-2}
  &&\hskip-7mm \big(b^{(1)}(t),f^{c}\big)=\sum\limits_{s=0}^{\infty}\,\frac{1}{s!}\,\mathrm{Tr}_{1,\ldots,s}
     \,b_{s}^{(1)}(t,1,\ldots,s)g_{s}(1,\ldots,s)\prod_{i=1}^{s}f_1^{0}(i)=\\
  &&\hskip+7mm=\mathrm{Tr}_{1}\,b_{1}^{0}(1)f_{1}(t,1),\nonumber
\end{eqnarray*}
where the operator $b_{s}^{(1)}(t)$ is determined by expansion (\ref{itvad}) and the one-particle
(marginal) density operator $f_{1}(t,1)$ is represented by the series expansion
\begin{eqnarray}\label{viterc}
   &&\hskip-12mm f_{1}(t,1)=\sum\limits_{n=0}^{\infty}\,\int\limits_0^tdt_{1}\ldots\int\limits_0^{t_{n-1}}dt_{n}\,
        \mathrm{Tr}_{2,\ldots,n+1}\mathcal{G}_{1}^{\ast}(t-t_{1},1)\mathcal{N}_{\mathrm{int}}^{\ast}(1,2)
        \prod\limits_{j_1=1}^{2}\mathcal{G}_{1}^{\ast}(t_{1}-t_{2},j_1)\ldots\\
   &&\hskip-5mm \times\prod\limits_{i_{n}=1}^{n}\mathcal{G}_{1}^{\ast}(t_{n}-t_{n},i_{n})
        \sum\limits_{k_{n}=1}^{n}\mathcal{N}_{\mathrm{int}}^{\ast}(k_{n},n+1)
        \prod\limits_{j_n=1}^{n+1}\mathcal{G}_{1}^{\ast}(t_{n},j_n)g_{1+n}(1,\ldots,n+1)\prod\limits_{i=1}^{n+1}f_1^0(i).\nonumber
\end{eqnarray}
In series (\ref{viterc}) the operator
$\mathcal{N}_{\mathrm{int}}^{\ast}(j_1,j_2)f_n=-\mathcal{N}_{\mathrm{int}}(j_1,j_2)f_n$
is an adjoint operator to operator (\ref{infOper1}) and the group
$\mathcal{G}_{1}^{\ast}(t,i)=\mathcal{G}_{1}(-t,i)$ is dual to group (\ref{grG})
in the sense of functional (\ref{averageb}).
For bounded interaction potentials series (\ref{viterc}) is norm convergent on the space
$\mathfrak{L}^{1}(\mathcal{H})$ under the condition that $t<t_{0}\equiv\big(2\,\|\Phi\|_{\mathfrak{L}(\mathcal{H}_{2})}
\|f_1^0\|_{\mathfrak{L}^{1}(\mathcal{H})}\big)^{-1}$.

The operator $f_{1}(t)$ represented by series (\ref{viterc}) is a solution of the Cauchy problem
of the quantum  Vlasov-type kinetic equation with initial correlations:
\begin{eqnarray}\label{Vls}
  &&\hskip-5mm\frac{\partial}{\partial t}f_{1}(t,1)=\mathcal{N}^{\ast}(1)f_{1}(t,1)+\\
  &&\hskip+5mm+\mathrm{Tr}_{2}\,\mathcal{N}_{\mathrm{int}}^{\ast}(1,2)
     \prod_{i_1=1}^{2}\mathcal{G}_{1}^{\ast}(t,i_1)g_{2}(1,2)
     \prod_{i_2=1}^{2}(\mathcal{G}_{1}^{\ast})^{-1}(t,i_2)f_{1}(t,1)f_{1}(t,2),\nonumber\\
     \nonumber\\
\label{Vlasov2c}
  &&\hskip-5mmf_{1}(t)|_{t=0}=f_{1}^0,
\end{eqnarray}
where the operator $\mathcal{N}^{\ast}(1)=-\mathcal{N}(1)$ is an adjoint operator
to operator (\ref{0}) in the sense of functional (\ref{averageb}) and the group
$(\mathcal{G}_{1}^{\ast})^{-1}(t)=\mathcal{G}_{1}^{\ast}(-t)=\mathcal{G}_{1}(t)$
is inverse to the group $(\mathcal{G}_{1}^{\ast})(t)$. This fact is proved similarly
as in case of a solution of the quantum BBGKY hierarchy represented by the iteration
series \cite{G12} (see also \cite{Pe71},\cite{Pe95}).

Thus, in case of initial states specified by one-particle (marginal) density operator 
(\ref{lins}) we establish that the dual quantum Vlasov hierarchy (\ref{vdh}) for additive-type 
marginal observables describes the evolution of quantum large particle system just as the quantum
Vlasov-type kinetic equation with initial correlations (\ref{Vls}).

\subsection{The mean field evolution of initial correlations}
The property of the propagation of initial correlations is a consequence of the validity
of the following equality for the mean value functionals of the limit $k$-ary marginal
observables in case of $k\geq2$
\begin{eqnarray}\label{dchaos}
    &&\hskip-12mm \big(b^{(k)}(t),f^{c}\big)=\sum\limits_{s=0}^{\infty}\,\frac{1}{s!}\,
       \mathrm{Tr}_{1,\ldots,s}\,b_{s}^{(k)}(t,1,\ldots,s)
       g_{s}(1,\ldots,s)\prod \limits_{j=1}^{s} f_1^0(j)=\\
    && =\frac{1}{k!}\mathrm{Tr}_{1,\ldots,k}\,b_{k}^{0}(1,\ldots,k)
       \prod _{i_1=1}^{k}\mathcal{G}_{1}^{\ast}(t,i_1)g_{k}(1,\ldots,k)
       \prod_{i_2=1}^{k}\mathcal{G}_{1}^{\ast}(-t,i_2)\prod\limits_{j=1}^{k}f_{1}(t,j),\quad k\geq2,\nonumber
\end{eqnarray}
where the limit one-particle (marginal) density operator $f_{1}(t,j)$ is represented
by series expansion (\ref{viterc}) and therefore it is governed by the Cauchy problem of the
quantum Vlasov-type kinetic equation with initial correlations (\ref{Vls}),(\ref{Vlasov2c}).

This fact is proved similarly to the proof of a property on the propagation of initial chaos
in a mean field scaling limit \cite{G11}.

Thus, in case of the limit $k$-ary marginal observables a solution of the dual quantum Vlasov
hierarchy (\ref{vdh}) is equivalent to a property of the propagation of initial correlations
for the $k$-particle marginal density operator in the sense of equality (\ref{dchaos}) or in
other words the mean field scaling dynamics does not create correlations.

We remark that the general approaches to the description of the evolution of states of quantum
many-particle systems within the framework of correlation operators and marginal correlation
operators were given in papers \cite{GerShJ},\cite{GP11} and \cite{GP}, respectively (see also
review \cite{G12}).

\section{On relationships of dynamics of observables and kinetic evolution of states}

We consider the relationships of dynamics of quantum many-particle systems described in terms
of marginal observables and dynamics described within the framework of a one-particle (marginal)
density operator governed by the quantum kinetic equation in the presence of initial correlations
in the general case, i.e. without any approximations like scaling limits as above in Section 2.
If initial states is completely specified by a one-particle (marginal) density operator,
using a non-perturbative solution of the dual quantum BBGKY hierarchy we prove that all possible
states at arbitrary moment of time can be described within the framework of a one-particle
density operator governed by the generalized quantum kinetic equation with initial correlations.

\subsection{Quantum dynamics of initial states specified by the one-particle density operator and correlations}
In case of initial states specified by sequence \eqref{ins}
the average values (mean values) of marginal observables \eqref{sdh} are defined by the positive
continuous linear functional on the space $\mathfrak{L}(\mathcal{F}_\mathcal{H})$
\begin{eqnarray}\label{averageB}
     &&\hskip-8mm \big(B(t),F^{c}\big)\doteq\sum\limits_{s=0}^{\infty}\frac{1}{s!}
         \,\mathrm{Tr}_{1,\ldots,s}\,B_{s}(t,1,\ldots,s)\,
         g_{s}^{\epsilon}(1,\ldots,s)\prod_{i=1}^{s}F_1^{0,\epsilon}(i).
\end{eqnarray}
For $F_1^{0,\varepsilon}\in{\mathfrak{L}^{1}(\mathcal{H})}$ and
$B_{s}^{0,\epsilon}\in \mathfrak{L}(\mathcal{H}_{s})$
series (\ref{averageB}) exists under the condition that
$\|F_1^{0,\varepsilon}\|_{\mathfrak{L}^{1}(\mathcal{H})}<e^{-1}$.

For mean value functional \eqref{averageB} the following representation holds
\begin{eqnarray}\label{w}
    &&\big(B(t),F^{c}\big)=\big(B(0),F(t\mid F_{1}(t))\big),
\end{eqnarray}
where $B(0)=(B_0,B_{1}^{0,\epsilon}(1),\ldots,B_{s}^{0,\epsilon}(1,\ldots,s),\ldots)\in
\mathfrak{L}_{\gamma}(\mathcal{F}_\mathcal{H})$ is a sequence of initial marginal observables,
and $F(t\mid F_{1}(t))=(1,F_1(t),F_2(t\mid F_{1}(t)),\ldots,F_s(t\mid F_{1}(t)),\ldots)$ is
a sequence of explicitly defined marginal functionals $F_s(t\mid F_{1}(t)),\,s\geq2$, with
respect to the following one-particle (marginal) density operator
\begin{eqnarray}\label{ske}
   &&\hskip-12mm F_{1}(t,1)= \sum\limits_{n=0}^{\infty}\frac{1}{n!}\,\mathrm{Tr}_{2,\ldots,{1+n}}\,\,
      \mathfrak{A}_{1+n}^{\ast}(t)g_{n+1}^{\epsilon}(1,\ldots,n+1)\prod_{i=1}^{n+1}F_{1}^{0,\epsilon}(i).
\end{eqnarray}
The generating operator $\mathfrak{A}_{1+n}^{\ast}(t)\equiv\mathfrak{A}_{1+n}^{\ast}(t,1,\ldots,n+1)$
of series expansion \eqref{ske} is the $(1+n)th$-order cumulant of groups of operators
$\mathcal{G}_n^{\ast}(t),\,n\geq1,$ dual to groups \eqref{grG} in the sense of functional \eqref{averageB},
namely
\begin{eqnarray*}\label{cu}
   &&\hskip-7mm \mathfrak{A}_{1+n}^{\ast}(t,1,\ldots,n+1)\doteq
       \sum\limits_{\mathrm{P}:\,(1,\ldots,n+1)={\bigcup}_i X_i}
       (-1)^{\mathrm{|P|}-1}({\mathrm{|P|}-1})!
       \prod_{X_i\subset\mathrm{P}}\mathcal{G}_{|X_i|}^{\ast}(t,X_i),
\end{eqnarray*}
where the symbol ${\sum}_\mathrm{P}$ means the sum over all possible partitions $\mathrm{P}$
of the set $(1,\ldots,n+1)$ into $|\mathrm{P}|$ nonempty mutually disjoint subsets
$ X_i\subset(1,\ldots,n+1)$.

The marginal functionals of the state $F_s(t\mid F_{1}(t)),\,s\geq2$, are represented by the following
series expansions:
\begin{eqnarray}\label{f}
     &&\hskip-12mm F_{s}\big(t,Y\mid F_{1}(t)\big)\doteq
        \sum _{n=0}^{\infty }\frac{1}{n!}\,\mathrm{Tr}_{s+1,\ldots,{s+n}}\,
        \mathfrak{G}_{1+n}\big(t,\{Y\},X\setminus Y\big)\prod_{i=1}^{s+n}F_{1}(t,i),
\end{eqnarray}
where we denote: $Y\equiv (1,\ldots,s),\,X\setminus Y\equiv (s+1,\ldots,s+n)$, and the
$(1+n)th$-order generating operator $\mathfrak{G}_{1+n}(t),\,n\geq0$, of this series
is determined by the following expansion
\begin{eqnarray}\label{skrrc}
   &&\hskip-7mm\mathfrak{G}_{1+n}(t,\{Y\},X\setminus Y)\doteq n!\,\sum_{k=0}^{n}\,(-1)^k\,\sum_{n_1=1}^{n}\ldots
       \sum_{n_k=1}^{n-n_1-\ldots-n_{k-1}}\frac{1}{(n-n_1-\ldots-n_k)!}\\
   &&\times\breve{\mathfrak{A}}_{1+n-n_1-\ldots-n_k}(t,\{Y\},s+1,\ldots,
       s+n-n_1-\ldots-n_k)\nonumber\\
   &&\times\prod_{j=1}^k\,\sum\limits_{\mbox{\scriptsize$\begin{array}{c}
       \mathrm{D}_{j}:Z_j=\bigcup_{l_j}X_{l_j},\\
       |\mathrm{D}_{j}|\leq s+n-n_1-\dots-n_j\end{array}$}}\frac{1}{|\mathrm{D}_{j}|!}
       \sum_{i_1\neq\ldots\neq i_{|\mathrm{D}_{j}|}=1}^{s+n-n_1-\ldots-n_j}\,\,
       \prod_{X_{l_j}\subset \mathrm{D}_{j}}\,\frac{1}{|X_{l_j}|!}\,\,
       \breve{\mathfrak{A}}_{1+|X_{l_j}|}(t,i_{l_j},X_{l_j}).\nonumber
\end{eqnarray}
In formula \eqref{skrrc} we denote by $\sum_{\mathrm{D}_{j}:Z_j=\bigcup_{l_j} X_{l_j}}$ the sum over all possible
dissections of the linearly ordered set $Z_j\equiv(s+n-n_1-\ldots-n_j+1,\ldots,s+n-n_1-\ldots-n_{j-1})$ on
no more than $s+n-n_1-\ldots-n_j$ linearly ordered subsets and we introduced the $(1+n)th$-order scattering cumulants
\begin{eqnarray*}
   &&\breve{\mathfrak{A}}_{1+n}(t,\{Y\},X\setminus Y)\doteq
       \mathfrak{A}_{1+n}^{\ast}(t,\{Y\},X\setminus Y)g_{s+n}^{\epsilon}(\theta(\{Y\}),X\setminus Y)
       \prod_{i=1}^{s+n}(\mathfrak{A}_{1}^{\ast})^{-1}(t,i),
\end{eqnarray*}
where the operator $g_{s+n}^{\epsilon}(\theta(\{Y\}),X\setminus Y)$ is specified initial correlations \eqref{lins},
the operator $(\mathfrak{A}_{1}^{\ast})^{-1}(t)$ is inverse to the operator $\mathfrak{A}_{1}^{\ast}(t)$
and it is used notations accepted above. We give examples of the scattering cumulants
\begin{eqnarray*}
   &&\hskip-5mm\mathfrak{G}_{1}(t,\{Y\})=\breve{\mathfrak{A}}_{1}(t,\{Y\})\doteq
      \mathfrak{A}_{1}^{\ast}(t,\{Y\})g_{s}^{\epsilon}(\theta(\{Y\}))\prod_{i=1}^{s}(\mathfrak{A}_{1}^{\ast})^{-1}(t,i),\\
   &&\hskip-5mm\mathfrak{G}_{2}(t,\{Y\},s+1)=
      \mathfrak{A}_{2}^{\ast}(t,\{Y\},s+1)g_{s+1}^{\epsilon}(\theta(\{Y\}),s+1)\prod_{i=1}^{s+1}(\mathfrak{A}_{1}^{\ast})^{-1}(t,i)-\\
   &&-\mathfrak{A}_{1}^{\ast}(t,\{Y\})g_{s}^{\epsilon}(\theta(\{Y\}))\prod_{i=1}^{s}(\mathfrak{A}_{1}^{\ast})^{-1}(t,i)
      \sum_{i=1}^s\mathfrak{A}_{2}^{\ast}(t,i,s+1)g_{2}^{\epsilon}(i,s+1)(\mathfrak{A}_{1}^{\ast})^{-1}(t,i)
      (\mathfrak{A}_{1}^{\ast})^{-1}(t,s+1).
\end{eqnarray*}
If $\|F_{1}(t)\|_{\mathfrak{L}^{1}(\mathcal{H})}<e^{-(3s+2)}$, then for arbitrary $t\in \mathbb{R}$
series expansion (\ref{w}) converges in the norm of the space $\mathfrak{L}^{1}(\mathcal{H}_{s})$ \cite{G12}.

We emphasize that marginal functionals of the state (\ref{f}) characterize the correlations generated
by dynamics of quantum many-particle systems in the presence of initial correlations.

\subsection{On an equivalence of mean value functional representations}
We prove the validity of equality (\ref{w}) for mean value functional (\ref{averageB}).

In a particular case of initial data specified by the additive-type marginal observables,
i.e. $B^{(1)}(0)=(0,B_{1}^{0,\epsilon}(1),0,\ldots)$, equality (\ref{w}) takes the form
\begin{eqnarray}\label{avmar-11}
   &&\big(B^{(1)}(t),F^{c}\big)=\mathrm{Tr}_{1}\,B_{1}^{0,\epsilon}(1)F_{1}(t,1),
\end{eqnarray}
where the one-particle (marginal) density operator $F_{1}(t)$ is determined by series expansion (\ref{ske}).
The validity of this equality is a result of the direct transformation of the generating operators of
expansions (\ref{af}) to adjoint operators in the sense of the functional (\ref{averageB}).

In case of initial data specified by the $s$-ary marginal observables
i.e. $B^{(s)}(0)=(0,\ldots,0,$ $B_{s}^{0,\epsilon}(1,\ldots,s),0,\ldots),\,s\geq2$, equality (\ref{w})
takes the following form:
\begin{eqnarray}\label{avmar-s}
    &&\hskip-12mm \big(B^{(s)}(t),F^{c}\big)=\frac{1}{s!}\mathrm{Tr}_{1,\ldots,s}\,
       B_{s}^{0,\epsilon}(1,\ldots,s)F_{s}\big(t,1,\ldots,s\mid F_{1}(t)\big),
\end{eqnarray}
where the marginal functional of the state $F_{s}(t\mid F_{1}(t))$ is represented
by series expansion (\ref{f}).

The proof of equality (\ref{avmar-s}) is based on the application of cluster expansions
to generating operators (\ref{cumulant}) of expansions (\ref{sdh}) which is dual to the
kinetic cluster expansions introduced in paper \cite{GT}. Then the adjoint series expansion
can be expressed in terms of one-particle (marginal) density operator (\ref{ske}) in the
form of the functional from the right-hand side of equality (\ref{avmar-s}).

In case of the general type of marginal observables the validity of equality (\ref{w})
is proven in much the same way as the validity of particular equalities (\ref{avmar-11})
and (\ref{avmar-s}).

\subsection{The generalized quantum kinetic equation with initial correlations}
As a result of the differentiation over the time variable of operator represented by series
(\ref{ske}) in the sense of the norm convergence of the space $\mathfrak{L}^{1}(\mathcal{H})$,
then the application of the kinetic cluster expansions \cite{GT},\cite{GTsm} to the generating
operators of obtained series expansion, for the one-particle (marginal) density operator we derive
the following identity
\begin{eqnarray}\label{gkec}
   &&\hskip-15mm\frac{\partial}{\partial t}F_{1}(t,1)=\mathcal{N}^{\ast}(1)F_{1}(t,1)+
      \epsilon\,\mathrm{Tr}_{2}\,\mathcal{N}_{\mathrm{int}}^{\ast}(1,2)F_{2}(t,1,2\mid F_{1}(t)),
\end{eqnarray}
where the operators $\mathcal{N}^{\ast}(1)=-\mathcal{N}(1)$ and
$\mathcal{N}_{\mathrm{int}}^{\ast}(1,2)=-\mathcal{N}_{\mathrm{int}}(1,2)$ are adjoint operators
in the sense of functional (\ref{averageb}) to operators (\ref{0}) and (\ref{int}), respectively,
and the collision integral is determined by series expansion (\ref{f}) for the marginal functional
of the state in case of $s=2$.
This identity we treat as the non-Markovian quantum kinetic equation. We refer to this evolution
equation as the generalized quantum kinetic equation with initial correlations.

We emphasize that the coefficients in an expansion of the collision integral of kinetic
equation \eqref{gkec} are determined by the operators specified initial correlations \eqref{ins}.
We remark also that in case of a system of particles with a $n$-body interaction potential the
collision integral of the corresponding quantum kinetic equation is determined by the marginal
functional of the state (\ref{f}) in case of $s=n$ \cite{G12}.

For the generalized quantum kinetic equation with initial correlations \eqref{gkec}
on the space $\mathfrak{L}^{1}(\mathcal{H})$ the following statement is true.

If $\|F_1^{0,\epsilon}\|_{\mathfrak{L}^{1}(\mathcal{H})}<(e(1+e^{9}))^{-1}$, the global
in time solution of initial-value problem of kinetic equation \eqref{gkec} is determined
by series expansion \eqref{ske}. For initial data
$F_1^{0,\epsilon}\in\mathfrak{L}^{1}_{0}(\mathcal{H})$ it is a strong (classical) solution
and for an arbitrary initial data it is a weak (generalized) solution.

We note that for initial data \eqref{ins} specified by a one-particle (marginal) density
operator, the evolution of states described within the framework of a one-particle (marginal)
density operator governed by the generalized quantum kinetic equation with initial correlations
\eqref{gkec} is dual to the dual quantum BBGKY hierarchy for additive-type marginal observables
with respect to bilinear form \eqref{averageB}, and it is completely equivalent to the description
of states in terms of marginal density operators governed by the quantum BBGKY hierarchy.

Thus, the evolution of quantum many-particle systems described in terms of marginal observables
can be also described within the framework of a one-particle (marginal) density operator governed
by the generalized quantum kinetic equation with initial correlations \eqref{gkec}.

\section{The asymptotic behavior of the generalized quantum kinetic equation with initial correlations}

We construct a mean field asymptotics of a solution of the generalized quantum kinetic
equation with initial correlations (\ref{gkec}). This asymptotics is governed by the quantum
Vlasov-type kinetic equation with initial correlations \eqref{Vls} derived above from the
dual quantum Vlasov hierarchy \eqref{vdh} for the limit marginal observables. Moreover,
a mean field asymptotic behavior of marginal functionals of the state (\ref{f}) describes
the propagation in time of initial correlations like established property \eqref{dchaos}.

\subsection{The limit theorem}
For solution (\ref{ske}) of the generalized quantum kinetic equation with initial correlations
(\ref{gkec}) the following mean field limit theorem is true \cite{GTsm}.
\begin{theorem}
If for the initial one-particle density operator $F_{1}^{0,\epsilon}\in\mathfrak{L}^{1}(\mathcal{H})$
exists the following limit:
$\lim_{\epsilon\rightarrow 0}\|\epsilon\,F_{1}^{0,\epsilon}-f_{1}^0\|_{\mathfrak{L}^{1}(\mathcal{H})}=0$
and $\lim_{\epsilon\rightarrow 0}\big\|g_{n}^{\epsilon}-g_{n}\big\|_{\mathfrak{L}^{1}(\mathcal{H}_n)}=0,$
then for $t\in(-t_{0},t_{0}),$ where
$t_{0}\equiv\big(2\,\|\Phi\|_{\mathfrak{L}(\mathcal{H}_{2})}\|f_1^0\|_{\mathfrak{L}^{1}(\mathcal{H})}\big)^{-1},$
there exists the mean field limit of solution (\ref{ske}) of the Cauchy problem of the generalized
quantum kinetic equation with initial correlations (\ref{gkec})
\begin{eqnarray}\label{1limc}
    &&\lim\limits_{\epsilon\rightarrow 0}\big\|\epsilon\,F_{1}(t)-
       f_{1}(t)\big\|_{\mathfrak{L}^{1}(\mathcal{H})}=0,
\end{eqnarray}
where the operator $f_{1}(t)$ is represented by series expansion (\ref{viterc}) and it is a solution
of the Cauchy problem of the quantum Vlasov-type kinetic equation with initial correlations \eqref{Vls},\eqref{Vlasov2c}.
\end{theorem}

The proof of this theorem is based on formulas of asymptotically perturbed cumulants
of groups of operators $\mathcal{G}_n^{\ast}(t),\,n\geq1,$ adjoint to groups \eqref{grG} in
the sense of functional \eqref{averageB}. Indeed, in a mean field limit for generating
evolution operators (\ref{skrrc}) of series expansion \eqref{f} the following equalities are valid:
\begin{eqnarray}\label{limc}
  &&\hskip-5mm\lim\limits_{\epsilon\rightarrow 0}\big\|\frac{1}{\epsilon^n}
     \mathfrak{G}_{1+n}(t,\{Y\},X\setminus Y)f_{s+n}\big\|_{\mathfrak{L}^{1}(\mathcal{H}_{s+n})}=0, \quad n\geq1,
\end{eqnarray}
and in case of the first-order generating evolution operator we have
\begin{eqnarray}\label{limc1}
  &&\hskip-8mm\lim\limits_{\epsilon\rightarrow 0}\big\|\big(\mathfrak{G}_{1}(t,\{Y\})-
     \prod_{j_1=1}^{s}\mathcal{G}_{1}^{\ast}(t,j_1)g_{s}(1,\ldots,s)
     \prod_{j_2=1}^{s}\mathcal{G}_{1}^{\ast}(-t,j_2)\big)f_{s}\big\|_{\mathfrak{L}^{1}(\mathcal{H}_{s})}=0,
\end{eqnarray}
respectively.

In view that under the condition $t<t_{0}\equiv(2\,\|\Phi\|_{\mathfrak{L}(\mathcal{H}_{2})}
\|\epsilon\,F_{1}^{0,\epsilon}\|_{\mathfrak{L}^{1}(\mathcal{H})})^{-1}$, for a bounded interaction
potential the series for the operator $\epsilon\,F_1(t)$ is norm convergent, then for $t<t_0$ the
remainder of solution series (\ref{ske}) can be made arbitrary small for sufficient large
$n=n_0$ independently of $\epsilon$. Then, using stated above asymptotic formulas, for each integer
$n$ every term of this series converges term by term to the limit operator $f_{1}(t)$ which is
represented by series (\ref{viterc}).

As stated above the mean field scaling limit (\ref{viterc}) of solution (\ref{ske}) of the
generalized quantum kinetic equation in the presence of initial correlations is governed by
the quantum Vlasov-type kinetic equation with initial correlations (\ref{Vls}).

Thus, we derived the quantum Vlasov-type kinetic equation with initial correlations \eqref{Vls}
from the generalized quantum kinetic equation \eqref{gkec} in the mean field scaling limit.
It is the same as the kinetic equation derived from the dual quantum Vlasov hierarchy
\eqref{vdh} for the mean field limit marginal observables.

\subsection{A mean field asymptotic behavior of marginal functionals of the state}
As we noted above in Section~3  in case of initial data (\ref{ins}) the evolution of all
possible correlations of quantum many-particle systems is described by marginal
functionals of the state (\ref{f}).

Since solution (\ref{ske}) of initial-value problem of the generalized quantum kinetic
equation with initial correlations (\ref{gkec}) converges to solution (\ref{viterc}) of
initial-value problem of the quantum Vlasov-type kinetic equation with initial correlations
(\ref{Vls}) as (\ref{1limc}), and equalities (\ref{limc}) and (\ref{limc1}) hold, then for
a mean field asymptotic behavior of marginal functionals of the state (\ref{f}) the following
equalities are true:
\begin{eqnarray*}\label{lf}
  &&\hskip-8mm\lim\limits_{\epsilon\rightarrow 0}\big\|\epsilon^{s}F_{s}(t,1,\ldots,s\mid F_{1}(t))-
     \prod _{j_1=1}^{s}\mathcal{G}_{1}^{\ast}(t,j_1)g_{s}(1,\ldots,s)
     \prod_{j_2=1}^{s}\mathcal{G}_{1}^{\ast}(-t,j_2)
     \prod\limits_{k=1}^{s}f_{1}(t,k)\big\|_{\mathfrak{L}^{1}(\mathcal{H}_{s})}=0,\\
  &&\hskip-8mm  s\geq2.
\end{eqnarray*}
These equalities describe the propagation of initial correlations in time in the mean field scaling
approximation.


\section{Conclusion and outlook}

In the paper the concept of quantum kinetic equations in case of the kinetic evolution,
involving correlations of particle states at initial time, for instance, correlations
characterizing the condensed states, was considered.
Two approaches were developed with a view to this purpose. One approach based on the description
of the evolution of quantum many-particle systems within the framework of marginal observables.
Another method consists in the possibility in case of initial states specified by a one-particle
marginal density operator and correlation operators to describe the evolution of states within
the framework of a one-particle (marginal) density operator governed by the generalized quantum
kinetic equation with initial correlations.

In case of pure states the quantum Vlasov-type kinetic equation with initial correlations (\ref{Vls})
can be reduced to the Gross--Pitaevskii-type kinetic equation. Indeed, in this case the
one-particle density operator $f_{1}(t)=|\psi_{t}\rangle\langle\psi_{t}|$ is a one-dimensional
projector onto a unit vector $|\psi_{t}\rangle\in\mathcal{H}$ and its kernel has
the following form: $f_{1}(t,q,q')=\psi(t,q)\psi^{\ast}(t,q')$. Then, if we consider quantum
particles, interacting by the potential which kernel $\Phi(q)=\delta(q)$ is the Dirac measure,
from kinetic equation (\ref{Vls}) we derive the Gross--Pitaevskii-type kinetic equation
\begin{eqnarray*}\label{gpeqc}
  &&\hskip-8mm i\frac{\partial}{\partial t}\psi(t,q)=-\frac{1}{2}\Delta_{q}\psi(t,q)+
    \int d q'd q''\mathfrak{g}(t,q,q;q',q'')\psi(t,q'')\psi^{\ast}(t,q)\psi(t,q),
\end{eqnarray*}
where the coupling ratio $\mathfrak{g}(t,q,q;q',q'')$ of the collision integral
is the kernel of the scattering length operator
$\prod_{i_1=1}^{2}\mathcal{G}_{1}^\ast(t,i_1)g_{2}(1,2)\prod_{i_2=1}^{2}\mathcal{G}_{1}^\ast(-t,i_2)$.
If we consider a system of quantum particles without initial correlations, then this kinetic
equation is the cubic nonlinear Schr\"{o}dinger equation.

This paper deals with a quantum system of a non-fixed (i.e. arbitrary but finite) number of identical
(spinless) particles obeying Maxwell--Boltzmann statistics. The obtained results can be extended to
quantum systems of bosons or fermions \cite{GP11}.

We emphasize, that one of the advantages of the developed approach to the derivation of the quantum
Vlasov-type kinetic equation with initial correlations from underlying dynamics governed by the
generalized quantum kinetic equation with initial correlations enables to construct the higher-order
corrections to the mean field evolution of quantum large particle systems.


\addcontentsline{toc}{section}{References}
\renewcommand{\refname}{References}


\begin{thebibliography}{99}

\bibitem{Go13}
                  F.~Golse,
                            \emph{On the dynamics of large particle systems in the mean field limit}.
                            Preprint arXiv:1301.5494 [math.AP], (2013).
\bibitem{S-R}
                  L.~Saint-Raymond,
                            \emph{Kinetic models for superfluids: a review of mathematical results}.
                            C. R. Physique. \textbf{5}, (2004), 65–-75.
\bibitem{PP09}
                  F.~Pezzotti and M.~Pulvirenti,
                            \emph{Mean-field limit and semiclassical expansion of quantum particle system}.
                            Ann. Henri Poincar\'{e}. \textbf{10}, (2009), 145--187.
\bibitem{FL}
                  J.~Fr\"{o}hlich, S.~Graffi and S.~Schwarz,
                            \emph{Mean-field and classical limit of many-body Schr\"{o}dinger dynamics for bosons}.
                            Commun. Math. Phys. \textbf{271}, (2007), 681--697.
\bibitem{BC}
                  W.~Bao and Y.~Cai,
                            \emph{Mathematical theory and numerical methods for Bose-Einstein condensation}.
                            Kinet. Relat. Models, \textbf{6}, (1), (2013), 1--135.
\bibitem{ESch}
                  L.~Erd\"{o}s and B.~Schlein,
                            \emph{Quantum dynamics with mean field interactions: a new approach}.
                            J. Stat. Phys. \textbf{134}, (5), (2009), 859--870.
\bibitem{ESchY2}
                  L.~Erd\"{o}s, B.~Schlein and H.-T.~Yau,
                            \emph{Derivation of the cubic nonlinear Schr\"{o}dinger equation from quantum dynamics
                            of many-body systems}.
                            Invent. Math. \textbf{167}, (3), (2007), 515--614.
\bibitem{Froh2}
                  J.~Fr\"{o}hlich and A.~Knowles,
                            \emph{A microscopic derivation of the time-dependent Hartree--Fock equation with Coulomb two-body interaction}.
                            J. Stat. Phys. \textbf{145}, (1), (2011), 23--50.
\bibitem{P11}
                  P.~Pickl,
                            \emph{A simple derivation of mean field limits for quantum systems}.
                            Lett. Math. Phys. \textbf{97}, (2), (2011), 151–-164.
\bibitem{KSS}
                  K.~Kirkpatrick, B.~Schlein and G.~Staffilani,
                            \emph{Derivation of the two dimensional nonlinear Schr\"{o}dinger equation from many body quantum dynamics}.
                            Amer. J. Math. \textbf{133}, (2011), 91--130.
\bibitem{EShY10}
                  L.~Erd\"{o}s, B.~Schlein and H.-T.~Yau,
                            \emph{Derivation of the Gross--Pitaevskii equation for the
                            dynamics of Bose--Einstein condensate}. Ann. of Math. \textbf{172}, (2010), 291--370.
\bibitem{LSSY}
                  E.~Lieb, R.~Seiringer, J.P.~Solovej, J.~Yngvason,
                            \emph{The mathematics of the Bose gas and its condensation}.
                            Oberwolfach Seminars, \textbf{34}. Birkh\"{a}user Verlag, Basel, 2005.
\bibitem{M1}
                  A.~Michelangeli,
                            \emph{Role of scaling limits in the rigorous analysis of Bose--Einstein condensation}.
                            J. Math. Phys. \textbf{48}, (2007), 102102.
\bibitem{CP}
                  T.~Chen and N.~Pavlovic,
                            \emph{The quintic NLS as the mean field limit of a Boson gas with three-body interactions}.
                             J. Funct. Anal. \textbf{260}, (4), (2011), 959--997.
\bibitem{Ch}
                  X.~Chen,
                            \emph{Second order corrections to mean field evolution for weakly interacting bosons in the case
                            of three-body interactions}. Arch. Rational Mech. Anal. \textbf{203}, (2012), 455–-497.
\bibitem{Sp91}
                  H.~Spohn,
                            \emph{Large Scale Dynamics of Interacting Particles}.
                            Springer-Verlag, 1991.
\bibitem{CGP97}
                  C.~Cercignani, V.I.~Gerasimenko and D.Ya.~Petrina,
                            \emph{Many-Particle Dynamics and Kinetic Equations}.
                            Kluwer Acad. Publ., 1997.
\bibitem{GT}
                  V.I.~Gerasimenko and Zh.A.~Tsvir,
                            \emph{A description of the evolution of quantum states by means of the kinetic equation}.
                            J. Phys. A: Math. Theor. \textbf{43}, (48), (2010), 485203.
\bibitem{GerS}
                  V.I.~Gerasimenko and V.O.~Shtyk,
                            \emph{Initial-value problem of the Bogolyubov hierarchy for quantum systems of particles}.
                            Ukrain. Math. J. \textbf{58}, (9), (2006), 1175--1191.
\bibitem{G12}
                  V.I.~Gerasimenko,
                            \emph{Hierarchies of quantum evolution equations and dynamics of many-particle correlations}.
                            (In: Statistical Mechanics and Random Walks: Principles, Processes and Applications.
                             N.Y.: Nova Science Publ., Inc., 2012), 233--288.
\bibitem{GTrmp}
                  V.I.~Gerasimenko and Zh.A.~Tsvir,
                            \emph{Mean field asymptotics of generalized quantum kinetic equation}.
                            Reports on Math. Phys. \textbf{70}, (2), (2012), 135--147.
\bibitem{BG}
                  G.~Borgioli and V.I.~Gerasimenko,
                           \emph{Initial-value problem of the quantum dual BBGKY hierarchy}.
                           Nuovo Cimento. \textbf{33 C}, (1), (2010), 71--78.
\bibitem{G11}
                  V.I.~Gerasimenko,
                            \emph{Heisenberg picture of quantum kinetic evolution in mean-field limit}.
                            Kinet. Relat. Models, \textbf{4}, (1), (2011), 385--399.
\bibitem{BogLect}
                  M.M.~Bogolyubov,
                            \emph{Lectures on Quantum Statistics. Problems of Statistical Mechanics of Quantum Systems}.
                            Rad. Shkola, 1949 (in Ukrainian).
\bibitem{SKB}
                 D.~Semkat, D.~Kremp and M.~Bonitz,
                            \emph{Kadanoff-Baym equations with initial correlations}.
                            Phys. Rev. E \textbf{59}, (2), (1999), 1557--1562.
\bibitem{KB}
                 L.P.~Kadanoff and G.~Baym,
                            \emph{Quantum Statistical Mechanics}.
                             W.A. Benjamin, 1962.
\bibitem{Pe71}
                  D.Ya.~Petrina,
                            \emph{On solutions of the Bogolyubov kinetic equations. Quantum statistics}.
                            Theor. Math. Phys. \textbf{13}, (3), (1972), 391--405.
\bibitem{Pe95}
                  D.Ya.~Petrina,
                            \emph{Mathematical Foundations of Quantum Statistical Mechanics. Continuous Systems}.
                            Kluwer Acad. Publ., 1995.
\bibitem{GerShJ}
                  V.I.~Gerasimenko and V.O.~Shtyk,
                            \emph{Evolution of correlations of quantum many-particle systems}.
                            J. Stat. Mech. Theory Exp. \textbf{3}, (2008), P03007, 24p.
\bibitem{GP11}
                  V.I.~Gerasimenko and D.O.~Polishchuk,
                            \emph{Dynamics of correlations of Bose and Fermi particles}.
                            Math. Meth. Appl. Sci. \textbf{34}, (1), (2011), 76--93.
\bibitem{GP}
                  V.I.~Gerasimenko and D.O.~Polishchuk,
                            \emph{A nonperturbative solution of the nonlinear BBGKY hierarchy for marginal
                            correlation operators}.
                            Math. Meth. Appl. Sci. \textbf{36}, (17), (2013), 2311--2328.
\bibitem{GTsm}
                  V.I.~Gerasimenko and Zh.A.~Tsvir,
                            \emph{On quantum kinetic equations of many-particle systems in condensed states.}
                            Physica A: Stat. Mech. Appl., \textbf{391}, (24), (2012), 6362--6366.
\end{thebibliography}
\end{document}